\documentclass[a4paper]{article}
\usepackage{epsfig}

\begin{document}                                                                \begin{center}                   
{\LARGE A Neural Chaos Model of Multistable Perception}
\vspace{6mm} \\
{\large Natsuki Nagao$^1$, Haruhiko Nishimura$^1$ and Nobuyuki Matsui$^2$} 
\end{center}
$^1$Studies of Information Science, Hyogo University of Education, 942-1 Yashiro-cho, Hyogo 673-1494, Japan, e-mail: \{nagao, haru\}@life.hyogo-u.ac.jp\\
$^2$Department of Computer Engineering, Himeji Institute of Technology, Himeji-shi, Hyogo 671-2201, Japan, e-mail: matsui@comp.eng.himeji-tech.ac.jp

\begin{abstract}
We present a perception model of ambiguous patterns based on the chaotic neural network and investigate the characteristics through computer simulations. The results induced by the chaotic activity are similar to those of psychophysical experiments and it is difficult for the stochastic activity to reproduce them in the same simple framework. Our demonstration suggests functional usefulness of the chaotic activity in perceptual systems even at higher cognitive levels. The perceptual alternation may be an inherent feature built in the chaotic neuron assembly.
\end{abstract}
\vspace{2mm} 

\section{Introduction}  
Perceptual alternation phenomena of ambiguous figures have been studied for a long time. Figure-ground, perspective (depth) and semantic ambiguities are well known (As an overview, for example, see \cite{attnv} and \cite{haken}). Actually, when we view the Necker cube which is a classic example of perspective alternation, a part of the figure is perceived either as front or back of a cube and our perception switches between the two different interpretations as shown in Fig.1. In this circumstance the external stimulus is kept constant, but perception undergoes involuntary and random-like change. The measurements have been quantified in psychophysical experiments and it becomes evident that the times between such changes are approximately Gamma distributed \cite{bor,bor2,haken}.

\begin{figure}
\centerline{\epsfig{file=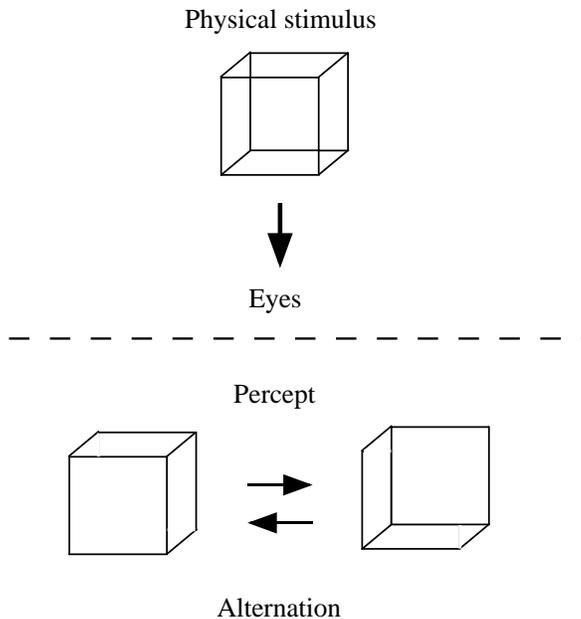,width=77mm}}
\caption{Perception of the Necker cube with its two alternative interpretations.}
\label{fig:3cube}
\end{figure}

Theoretical model approaches to explaining the facts have been made mainly from three situations based on the synergetics \cite{dit,dit2,dante}, the BSB (brain-state-in-a-box) neural network model \cite{kawamoto,riani,matsui2}, and the PDP (parallel distributed processing) schema model \cite{rume,sakai,ino}. Common to these approaches is that top-down designs are applied so that the model can be manipulable by a few parameters and upon this basis fluctuating sources are brought in. The major interests seem to be not in the relation between the whole function and its element (neuron), but in the model building at the phenomenological level.

Until now diverse types of chaos have been confirmed at several hierarchical levels in the real neural systems from single cells to cortical networks (e.g.~ionic channels, spike trains from cells, EEG) \cite{arbib}. This suggests that artificial neural networks based on the McCulloch-Pitts neuron model \cite{MC43} should be re-examined and re-developed. Chaos may play an essential role in the extended frame of the Hopfield neural network \cite{hopf} beyond the only equilibrium point attractors. To make this point clear, following the model of chaotic neural network \cite{aihara2}, the dynamic learning and retrieving features of the associative memory have been studied \cite{adachi,foo24}. In this paper, we present a perception model of ambiguous patterns based on the chaotic neural network from the viewpoint of bottom-up approach \cite{iconip97}, aiming at the functioning of chaos in dynamic perceptual processes.

\section{Model and Method} 
The chaotic neural network (CNN) composed of N chaotic neurons is described as \cite{aihara2,foo24}
\begin{eqnarray}
X_i(t+1) &=& f(\eta_i(t+1)+\zeta_i(t+1))~~,\\
\eta_i(t+1) &=& \sum_{j=1}^N w_{ij} \sum_{d=0}^{t} k_{f}^{d} X_{j}(t-d)~~,\\ 
\zeta_i(t+1) &=& - \alpha \sum_{d=0}^{t}k_{r}^{d} X_{i}(t-d) - \theta_{i}~~,
\end{eqnarray}
where $X_{i}$~:~output of neuron $i (-1 \leq X_i \leq 1),~w_{ij}$~:~synaptic weight from neuron $j$ to neuron $i,~\theta_{i}$~:~threshold of neuron $i,~k_{f}(k_{r}$)~:~decay factor for the feedback(refractoriness) $(0 \leq k_{f},~k_{r} <1),~\alpha$~:~refractory scaling parameter,~$f$~:~output function defined by $f(y)=tanh(y/2\varepsilon)$ with the steepness parameter $\varepsilon$. Owing to the exponentially decaying form of the past influence,~Eqs.(2) and (3) can be reduced to
\begin{eqnarray}
\eta_{i}(t+1) &=& k_{f} \eta_{i}(t) + \sum_{j=1}^{N} w_{ij} X_{j}(t) ~~,\\
\zeta_{i}(t+1) &=& k_{r} \zeta_{i}(t) - \alpha X_{i}(t) + a~~, 
\end{eqnarray}
where $a$ is temporally constant $a \equiv -\theta_{i}(1-k_{r})$. All neurons are updated in parallel, that is, synchronously. The network corresponds to the conventional discrete-time Hopfield network :
\begin{eqnarray}
X_i(t+1) = f\Bigl(\sum_{j=1}^N w_{ij} X_j(t) - \theta_i\Bigl)
\end{eqnarray}
when $\alpha = k_f = k_r = 0$~(Hopfield network point (HNP)). The asymptotical stability and chaos in discrete-time neural networks are theoretically investigated in Refs. \cite{M-W,C-A}.

Under external stimuli, Eq.(1) is influenced as
\begin{eqnarray}
X_{i}(t+1) = f \bigl( \eta_{i}(t+1) + \zeta_{i}(t+1) + \sigma_{i} \bigr),
\end{eqnarray}
where $\{ \sigma_{i} \}$ is the effective term by external stimuli. This is a simple and unartificial incorporation of stimuli as the changes of neural active potentials.

The two competitive interpretations are embedded in the network as minima of the energy map :
\begin{eqnarray}
E=-\frac{1}{2}\sum_{ij}w_{ij}X_iX_j
\end{eqnarray}
at HNP. This is done by using a iterative perception learning rule for $p(<N)$ patterns $\{ \xi_i^\mu\} \equiv (\xi_1^\mu,\cdots,\xi_N^\mu),(\mu=1,\cdots,p;\xi_i^\mu=~+1 or -1)$ in the form :
\begin{eqnarray}
w_{ij}^{new} = w_{ij}^{old} + \sum_\mu\delta w_{ij}^\mu
\end{eqnarray} 
with
\begin{eqnarray}
\delta w_{ij}^\mu=\frac{1}{N}\theta (1-\gamma_i^\mu)\xi_i^\mu\xi_j^\mu ,
\end{eqnarray}
where $\gamma_i^\mu \equiv \xi_i^\mu \sum_{j=1}^Nw_{ij}\xi_j^\mu$ and $\theta(h)$ is the unit step function. The learning mode is separated from the performance mode by Eq.(7).

The conceptual picture of our model is shown in Fig.2. Under the external stimulus $\{ \sigma_i \}$, chaotic activities arise on the neural network and cause the transitions between stable states of HNP. This situation corresponds to the dynamic multistable perception. Note that $\theta (1-\gamma_i^\mu)$ turns off learning for overlearned patterns. This will be empirically shown in the next section to have dynamics characteristic of a chaotic dynamical system.

\section{Simulations and Results}
To carry out computational experiments, we use the $12 \times 13 (N=156)$ non-orthogonal 10 random patterns $\{ \xi_i^\nu \}(\nu=1,\cdots,10,; i=1,\cdots,N)$ as a set of ambiguous figure stimuli: $\{\sigma_i\} = s\{\xi_{i}^{\nu}\}$. $s$ is the strength factor of stimulation. For each of them, two interpretation patterns $\{\xi_i^{\nu1}\}$ and $\{\xi_i^{\nu2}\}$ are prepared by changing 15 white ($\xi_i=-1$) pixels to black ($\xi_i=+1$) ones which do not overlap between $\nu1$ and $\nu2$ as shown by shaded and dotted in Fig.3, and are memorized following the above learning rule $(p=20)$.

\begin{figure}[t]
\centerline{\epsfig{file=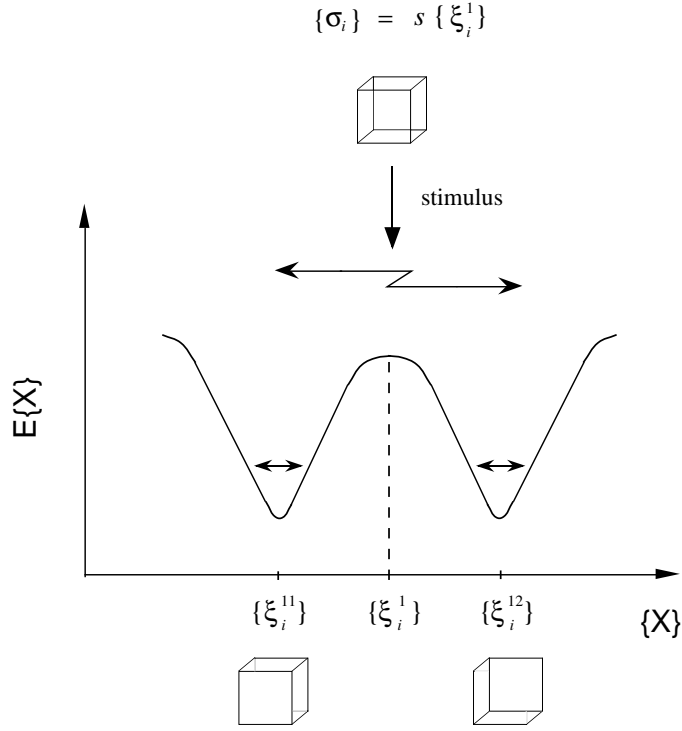,width=90mm}}
\caption{Conceptual picture illustrating state transitions induced by chaotic activity}
\label{fig:ncimage}
\end{figure}

Figure 4 shows a time series evolution of CNN $(k_f=0.5,k_r=0.8,\alpha=0.34,a=0,\varepsilon=0.015)$ under the stimulus $\{\sigma_i\}=0.7\{\xi_i^1\}$. Here,
\begin{eqnarray}
m^{11}(t)=\frac{1}{N}\sum_{i=1}^N\xi_i^{11}X_i(t)
\end{eqnarray}

\begin{figure}
\centerline{\epsfig{file=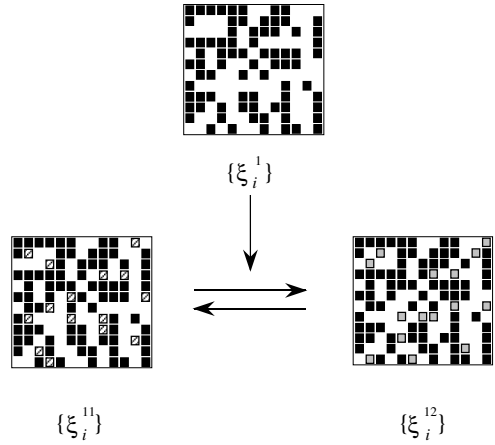,width=64mm}}
\caption{Pattern states of the neural network correspond to the ambiguous figure and its interpretations in Fig.1. White and other pixels represent the states -1 and +1, respectively.}
\label{fig:ncpat}
\end{figure}

\begin{figure}
\centerline{\epsfig{file=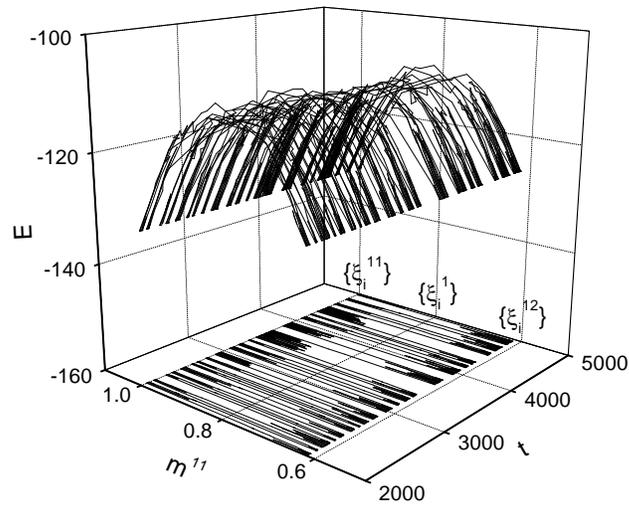,width=82mm}}
\caption{Time series of the overlap with $\{\xi_i^{11}\}$ and the energy map under the stimulus $\{\xi_i^1\}$.}
\label{fig:cnn3d}
\end{figure}
and is called the overlap of the network state $\{X_i\}$ and the interpretation pattern $\{\xi_i^{11}\}$. A switching phenomenon between $\{\xi_i^{11}\}$ ($m^{11}=1.0$) and $\{\xi_i^{12}\}$ ($m^{11}=0.62$) can be observed. Bursts of switching are interspersed with prolonged periods during which $\{X_i\}$ trembles near $\{\xi_i^{11}\}$ or $\{\xi_i^{12}\}$. Evaluating the maximum Lyapunov exponent \cite{foo21} to be positive ($\lambda_1=0.26$), we find that the network is dynamically in chaos. In the cases $\lambda_1<0$, such switching phenomena do not arise. 

\begin{figure}
\centerline{\epsfig{file=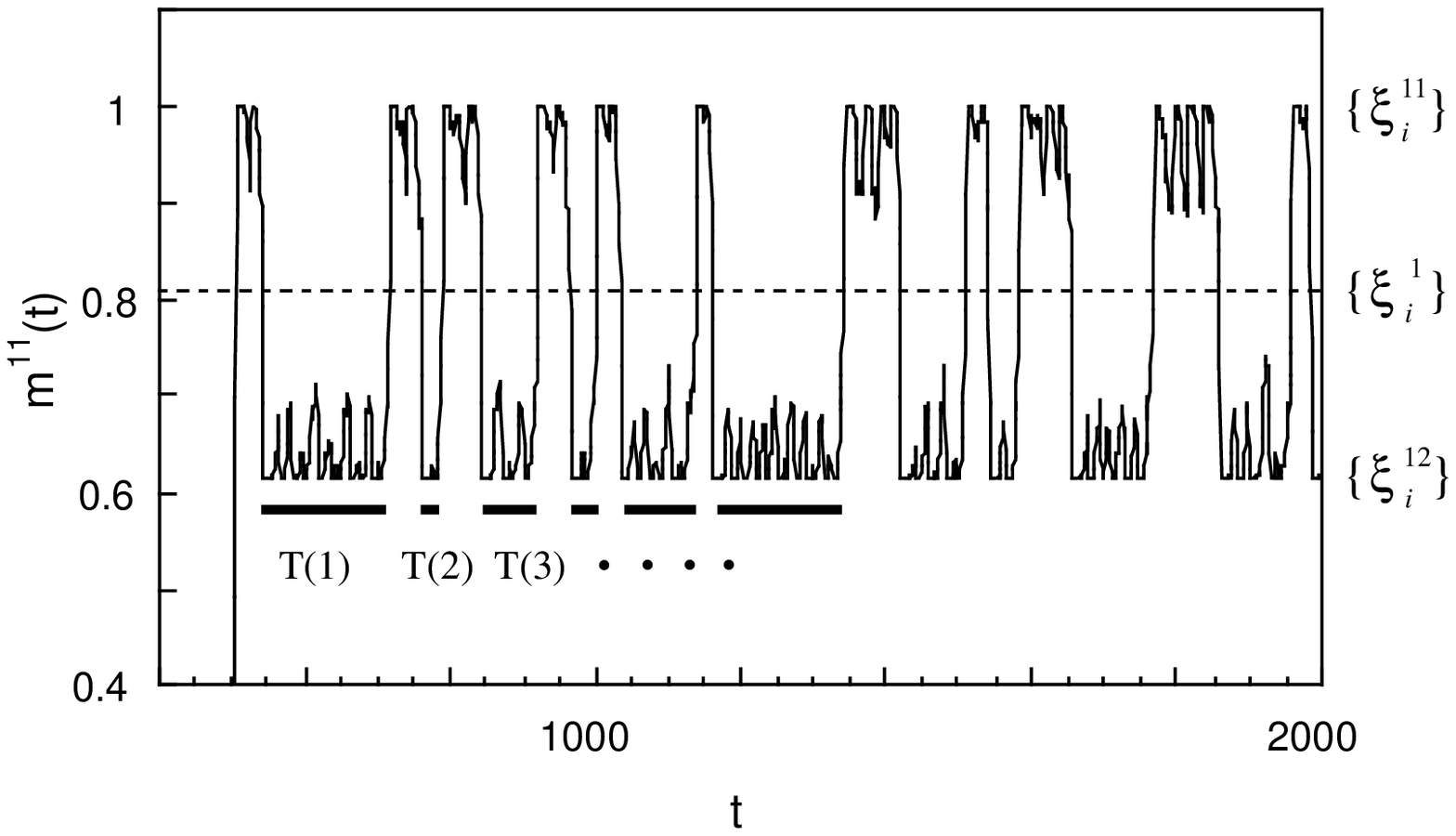,width=90mm}}
\caption{Time series of the overlap with $\{\xi_i^{11}\}$ under the stimulus $\{\xi_i^1\}$, magnified for t-axis.}
\label{fig:zoom3d}
\end{figure}

From the $2 \times 10^5$ iteration data (until $t=2 \times 10^5$) of Fig.4, we get 1257 events staying near one of the two interpretations, $\{\xi_i^{12}\}$. As can be seen from Fig.5 magnified for t-axis, they have various persistent durations $T(1) \sim T(1257)$ which seem to have a random time course by the return map in $(T(n),T(n+1))$ shown in Fig.6. From the evaluation of the autocorrelation function for $T(n)$, $C(k)=<T(n+k)T(n)>-<T(n+k)><T(n)>$ (here, $< >$ means an average over time), we get $-0.06<C(k)/C(0)<0.06$ against $k=1 \sim 100$. This suggests successive durations $T(n)$ are independent.
The frequency of occurrence of $T$ is plotted for 1257 events in Fig.7. The distribution is well fitted by Gamma distribution

\begin{figure}
\centerline{\epsfig{file=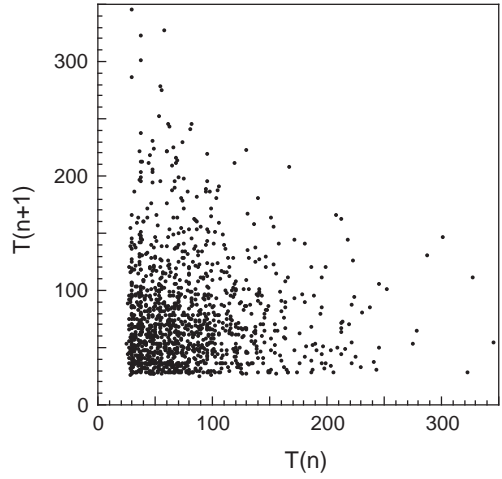,width=65mm}}
\caption{Return map of the persistent durations staying one of the two interpretations for the data ($T(1) \sim T(1257)$) in Fig.3 (up to $t=2 \times 10^5$).}
\label{fig:retmap}
\end{figure}

\begin{figure}
\centerline{\epsfig{file=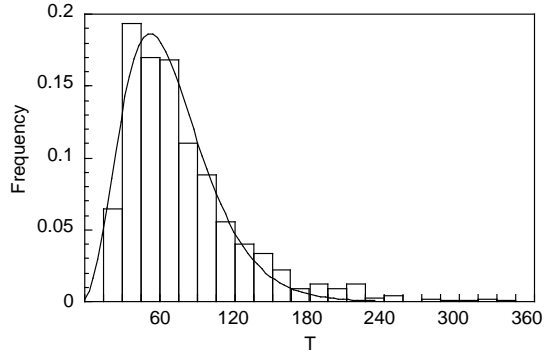,width=72mm}}
\caption{Frequency distribution of the persistent durations T(n) and the corresponding Gamma distribution.}
\label{fig:gamma}
\end{figure}

\begin{eqnarray}
G(\tilde{T})=\frac{b^n \tilde{T}^{n-1}e^{-b \tilde{T}}}{\Gamma(n)}
\end{eqnarray}
with $b=0.918, n=4.68(\chi^2=0.0033, r=0.98)$, where $\Gamma(n)$ is the Euler-Gamma function. 
$\tilde{T}$ is the normalized duration $T/15$ and here 15 step interval is applied to determine the relative frequencies.

The results are in good agreement with the characteristics of psychophysical experiments \cite{bor,bor2,haken}. Similar results to the above example are obtained in the appropriate parameter regions where the network may induce chaotic activities under external stimuli. It is found that aperiodic spontaneous switching does not necessitate some stochastic description as in the synergetic model \cite{dit,dit2}.

These results can not be easily explained through the use of a standard Hopfield network with a stochastic fluctuation forcing function. We look into the case that the stochastic fluctuation $\{ F_i \}$ is attached to Eq.(6) of HNP together with the external stimulus $\{ \sigma_i \}$ :

\begin{eqnarray}
X_i(t+1) = f\Bigl(\sum_{j=1}^{N} w_{ij} X_{j}(t) + \sigma_i + F_i(t)\Bigl) ~~~,
\end{eqnarray}
where
\begin{eqnarray}
\left\{
\begin{array}{ll}
<F_i(t)> = 0 \vspace{2mm}\\
<F_i(t)F_j(t')> = D^2 \delta_{tt'} \delta_{ij} ~~~~.
\end{array}
\right.
\end{eqnarray}
Using this equation in the same framework, we examined many cases with different values of the noise strength $D$, but couldn't find successive alternation phenomena. Figure 8 is a typical result in $s = 0.5, D = 0.65$ and is far from the realization of sudden perceptual changes. In such a simple scheme, the noise does not drive a quick motion of $\{X_i\}$ for the energy barrier. In Fig.9 the frequency of occurrence $T$ is plotted for 1387 events obtained until $t=2 \times 10^5$. The distribution becomes the exponential-like, not Gamma distribution.

\begin{figure}
\centerline{\epsfig{file=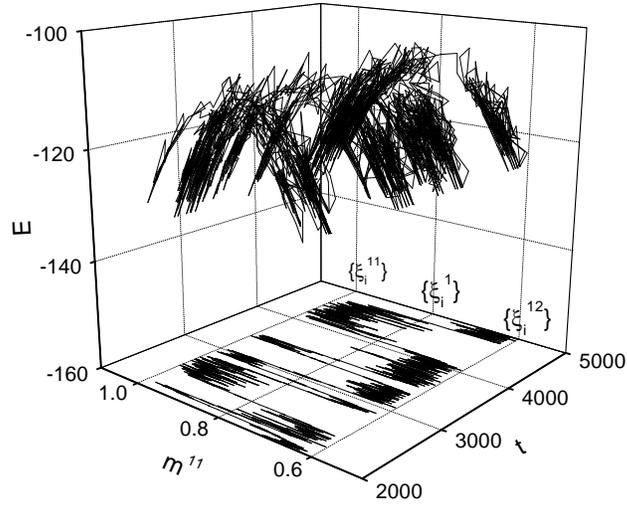,width=82mm}}
\caption{Time series of the overlap with $\{\xi_i^{11}\}$ and the energy map under the stimulus $\{\xi_i^1\}$. In the case of stochastic activity in Eq(13). (Compare with Fig.4.)}
\label{fig:snn3d}
\end{figure}

\begin{figure}
\centerline{\epsfig{file=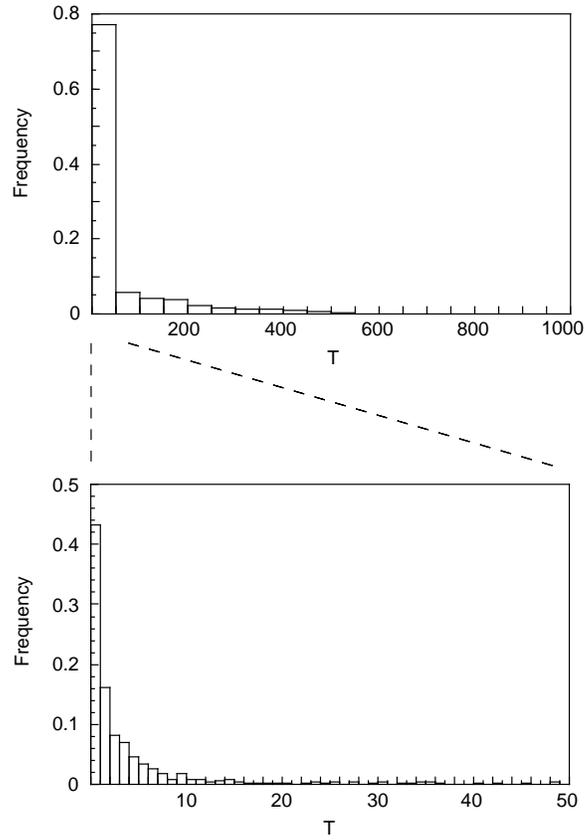,width=77mm}}
\caption{Frequency distribution in the case of stochastic fluctuations.}
\label{fig:distsnn}
\end{figure}

\section{Conclusion}
We have shown that the neural chaos leads to perceptual alternations as  responses to ambiguous stimuli in the chaotic neural network. Its emergence is based on the simple process in a realistic bottom-up framework. In the same stage, similar results can not be obtained by the stochastic activity. In order to compare the simulation results with experimental ones in a concrete form, there remain problems which are to analyze the relationship between the iteration (step) time and the visual real time, and to study the switching dependence on the cube's orientation or size.

Finally, our demonstration suggests functional usefulness of the chaotic activity in perceptual systems even at higher cognitive levels. The perceptual alternation appears to be an inherent feature built in the chaotic neuron assembly. It may be interesting to study the brain with the experimental technique (e.g., fMRI) under the circumstance where the perceptual alternation is running.

\end{document}